\documentclass[11pt]{article}
\usepackage{epsf,latexsym}
\newcommand{\pr}{{Phys.\ Rev.\/}}
\newcommand{\np}{{Nucl.\ Phys.\/}}
\newcommand{\pl}{{Phys.\ Lett.\/}}

\title{Investigating Confinement in \\ Dually Transformed $U(1)$ Lattice Gauge Theory
\thanks{Supported by Fonds zur F\"orderung der wissenschaftlichen Forschung, Proj.~P11156-PHY}}

\author{Martin Zach, Manfried Faber and Peter Skala  \\
Institut f\"ur Kernphysik, Technische Universit\"at Wien, \\ A-1040 Vienna, Austria}
       
\date{}

\begin{document}

\maketitle

\begin{abstract}
The dually transformed path integral of four-dimensional $U(1)$ lattice gauge theory is used for the calculation of expectation values in the presence of external charges. Applying the dual simulation to flux tubes for charge distances up to around 20 lattice spacings, we find a deviation from the behaviour of a dual type-II superconductor in the London limit at large distances. The roughening of the flux tube agrees with the effective string picture of confinement. Further we show that finite temperature effects are negligible for time extents larger than the charge distance. Finally we analyze the different contributions to the total energy of the electromagnetic field. The parallel component of the magnetic field turns out to have the strongest influence on the deviation from the linear behaviour at small charge distances.
\end{abstract}
\section{Introduction}
The mechanism of quark confinement within hadrons is still an open problem. Numerical simulations of QCD on  space-time lattices demonstrate the confinement phenomenon but could not yet clarify completely the mechanism for the formation of a gluonic flux tube. Effective theories have been created to improve our understanding of this mechanism. An intuitive physical picture is the dual superconductor model \cite{mandelstam}: Dynamically generated colour magnetic monopole currents form a solenoid which squeezes the chromoelectric flux between quarks into a narrow flux tube. The energy of this flux tube increases linearly with its length and confinement is achieved. Another approach is the effective string theory \cite{nambu, luescher} which should describe flux tubes at large quark distances. It is an interesting goal of lattice gauge theories to test the validity of such effective models and therefore deepen our understanding of the confinement mechanism.

$U(1)$ lattice gauge theory in four dimensions undergoes a phase transition at strong coupling and has been widely used as a prototype of a confining theory. In $U(1)$ theory the condensation of magnetic monopoles is made responsible for the confinement of electric charges. The results of lattice simulations have been interpreted in terms of the dual superconductor picture of confinement, e.g. the validity of a dual London relation has been checked \cite{haymaker}, and the $U(1)$ results for electric fields and monopole currents were found to agree with the predictions of a classical effective model considering also string fluctuations \cite{pl95}. Most of the obtained results are expected to hold also in non-Abelian gauge theories, at least qualitatively \cite{peter}. The data accumulated so far, however, for larger flux tube lengths are not of sufficient accuracy to perform a quantitative analysis in terms of effective models for the formation of a gluonic flux tube.

Furthermore, it was realized many years ago \cite{banks} that one can perform a duality transformation of the path integral in compact Abelian gauge theories. In this way a new partition function is obtained which can be regarded as a limit of the dual non-compact Abelian Higgs model \cite{froehlich}. We will use the duality relation for expectation values in the presence of external charges to clarify the connection between dually transformed $U(1)$ theory and the dual superconductor model. Besides this support for the interpretation of the results, the dual theory provides a very efficient tool for simulating flux tubes. This fact has already been realized in refs.~\cite{greensite, caselle}, but to our knowledge no dual simulations of four-dimensional $U(1)$ lattice gauge theory with external sources have been performed yet.

The organisation of this article is as follows: In section 2 we review the duality transformation of $U(1)$ lattice gauge theory for the Wilson action with special consideration of fields and their squares in the presence of external charges. The consequences for the interpretation in the dual superconductor picture are discussed by connecting the results to a dual Higgs model in section 3. After pointing out the advantages of dual simulations of flux tubes in section 4, we present our results for electric field strength and magnetic current in the presence of a charge--anticharge--pair in section 5, analyzing the flux tube as a function of spatial lattice size and temperature. In section 6 we return to the dual superconductor picture, subjecting our data to a quantitative comparison with the London model. In section 7 the dependence of the flux tube width on its length is investigated and compared with the prediction of an effective string theory. Finally in section 8 we calculate the total electromagnetic energy and analyze the various components contributing to it.

\section{The duality transformation of $U(1)$ lattice gauge theory}

We begin with the usual prescription for simulating $U(1)$ lattice gauge theory. Expectation values of physical observables ${\cal O}$ in the presence of a static charge pair at distance $d$ are determined by the correlation function
\begin{equation} \label{correlation}
\langle {\cal O}(x) \rangle_{Q\bar{Q}} = \frac{\langle L(0) L^+(d) {\cal O}(x) \rangle}{\langle L(0) L^+(d) \rangle} - \langle {\cal O} \rangle,
\end{equation}
where $L(\vec{r})$ is the Polyakov loop $L(\vec{r})=\prod_{k=1}^{N_{t}}U_{x=(\vec r,ka),\mu=0}$ and the angle brackets denote the evaluation of the path integral using the standard Wilson action
\begin{equation} \label{action}
S_W = \beta \sum_{x, \mu < \nu} [1-\cos (d \theta)_{x,\mu\nu}].
\end{equation}
$\beta=1/e^2$ is the inverse coupling, and $(d \theta)_{x,\mu\nu}$ is the discretized exterior derivative of the phases $\theta$ of the link variables $U_{x,\mu}=e^{i\theta_{x,\mu}}$ and is assigned to a plaquette of size $a^2$. Below we will use the notation of lattice differential forms and therefore suppress indices. The observables $\cal O$ we are interested in are the electric field strength, the magnetic current and the electric and magnetic energy densities. For the field strength $F$ we use the identification 
$a^2 e F = \sin (d \theta)$
which was shown to fulfil the electric Gauss law for the Wilson action \cite{pl95}. The magnetic current can be constructed via the dual Maxwell equations. The squared fields are calculated in the usual way by means of the operator $\beta \cos (d \theta)$.

Let us briefly review the most important steps of the duality transformation. We start from the $U(1)$ partition function using the Wilson action which reads \cite{varenna}
\begin{equation}
Z = \prod \int^{\pi}_{-\pi} d \theta \;\; \exp\{- \beta \sum
(1 - \cos d\theta)\},
\end{equation}
where the product is over all links on the lattice, and the sum in the exponent is over the plaquettes. Expanding the Boltzmann factor for each plaquette in a Fourier series one can write
\begin{equation}\label{ztp}
Z = \prod \int^{\pi}_{-\pi} d \theta \;\;
     \prod \sum_k \;\; \exp\{i(k,d\theta)\} e^{-\beta} I_{\parallel k \parallel}(\beta)\; \; = \; \; (2 \pi)^{4N}    \prod \sum_{k}
\;\; e^{-\beta} I_{\parallel k \parallel} (\beta) \biggr\vert_{\delta k = 0},
\end{equation}
where we have used the plaquette variables $k$ and the property of the inner product $(k, d \theta) = (\delta k, \theta)$, and then performed the integration over the links $\theta$. $N$ is the number of lattice sites, and the product has to be performed over all plaquettes. $k$ is an integer valued 2-form corresponding to the field strength $F$. The constraint $\delta k = 0$ describes the absence of electric charges, $\delta F = J_{e} = 0$. We solve this constraint by introducing an integer valued 3-form $l$ with
\begin{equation} 
k = \delta l,
\label{kl}
\end{equation}
\begin{equation}
Z =    (2 \pi)^{4N}    \prod \sum_{l}
\;\; e^{-\beta} I_{\parallel {\delta l} \parallel}(\beta) \;\; = \;\; (2 \pi)^{4N}  \prod \sum_{^*l}
\;\; e^{-\beta} I_{\parallel d {}^*l \parallel}(\beta),
\label{zdual}
\end{equation}
where we have finally switched to the dual link variables $^*l$, representing an integer valued dual potential. The product in the partition function extends over the links of the dual lattice.

The next step is the inclusion of external charges into the $U(1)$ path integral. ${\cal L_+}$ and ${\cal L_-}$ denote the world-lines of a static charge pair. Starting from 
\begin{equation}\label{wjein}
Z_{Q\bar{Q}} = \prod \int^{\pi}_{-\pi} d \theta \;\; \exp\{- \beta \sum
(1 - \cos d\theta)\} \prod_{\cal L_+} e^{-i \theta} \prod_{\cal L_-} 
e^{i \theta},
\end{equation}
one arrives at an expression equivalent to (\ref{ztp}), with a modified constraint $\delta k = \pm 1$ for links belonging to ${\cal L_{\pm}}$ (equivalent to $e\, \delta k = J_e$). This constraint is solved by an additional integer 2-form $n$:
\begin{equation}
k = \delta l + n, \;\; {\rm with} \;\; e\,\delta n = J_e, \;\;  \textrm{resp.} \;\; {}^*k = d \,{}^*l + {}^*n, \;\; {\rm with} \;\; e\,d \,{}^* n = {}^*J_e.
\end{equation}
External electric currents are represented by the boundaries of a dual Dirac sheet in the dual theory. The dually transformed path integral therefore reads
\begin{equation}\label{wje}
Z_{Q\bar{Q}} \;\; = \;\; (2 \pi)^{4N} \;\; \prod \sum_{^*l}
\;\; e^{-\beta} I_{\parallel d {}^*l + {}^*n \parallel}(\beta),
\end{equation}
where the product is performed over all links of the dual lattice. The dual Dirac sheet $^*n$ can be fixed as the minimal straight area between the two Polyakov loops, while the integer gauge field $^*l$ is kept as dynamical variable and updated in simulations. Fluctuations in $^*l$ result in fluctuations of the flux string $d \,{}^*l + {}^*n $.

Further, we will perform the duality transformation for the correlation function (\ref{correlation}). For the determination of the flux $F_{\Box_0}$ through a given plaquette ${\Box_0}$, we have to insert the operator $\sqrt{\beta} \sin d \theta_{\Box_0}$ into the path integral in addition to the Polyakov loops. So we can write
\begin{eqnarray}
<F_{\Box_0}>_{Q\bar{Q}} \;\; = \;\; Z^{-1}_{Q\bar{Q}} \;\; \Im \prod \int^{\pi}_{-\pi} d \theta \; \prod \sum_k \;  \prod_{\cal L_+} e^{-i \theta} \prod_{\cal L_-} 
e^{i \theta} \; \frac{\sqrt{\beta}}{2i} \left( \prod_{\Box_0} e^{i \theta} - 
\prod_{\Box_0} e^{-i \theta} \right) \times \nonumber \\
\times \;\; \exp\{i(\delta k,\theta)\} \; \;e^{-\beta} \;I_{\parallel k \parallel}(\beta) \; .
\end{eqnarray}
Taking the imaginary part reflects the change of sign of the field strength under an exchange of charge and anticharge, i.~e.~under complex conjugation. Our constraint for $\delta k$ is further modified: $\delta k$ has to be non-zero along the electric current and the boundary of the considered plaquette $\Box_0$. Solving this constraint by a redefinition of $k$ on this plaquette ($k \to k \pm 1$) and using the identity for modified Bessel functions
\begin{equation}
k \; I_k(\beta) = \frac{\beta}{2} \left[ I_{k-1}(\beta) - I_{k+1}(\beta) \right],
\end{equation}
one obtains
\begin{equation}\label{sin}
<F_{\Box_0}>_{Q\bar{Q}} \;\; = \;\; Z^{-1}_{Q\bar{Q}} \;\; (2 \pi)^{4N} \;\;\prod \sum_{^*l} \;\;\frac{1}{\sqrt{\beta}} {}^*k_{\Box_0} \;\; e^{-\beta} I_{\parallel d {}^*l + {}^*n \parallel}(\beta),
\end{equation} 
where $^*k = d \,{}^*l + {}^*n$ as before. Note that in this dual expression the validity of Gauss' law for electric charges is checked easily because of 
$d\, {}^*k/\sqrt{\beta} = \,{}^*J_e$.

Performing the duality transformation for the expectation value of $\beta \cos d \theta_{\Box_0}$ in an analogous way leads to
\begin{equation}\label{cos}
<F^2_{\Box_0}>_{Q\bar{Q}}\;\; = \;\; Z^{-1}_{Q\bar{Q}} \;\; (2 \pi)^{4N} \;\; 
\prod \sum_{^*l} \;\; \beta \frac{I^{\prime}_{{{}^*k}_{\Box_0}}(\beta)}{I_{{{}^*k}_{\Box_0}}(\beta)} \;\; e^{-\beta} I_{\parallel d {}^*l + {}^*n \parallel}(\beta) \;\; - \;\; <F^2>_{vac}.
\end{equation} 
We would like to mention that an identity equivalent to (\ref{cos}) was already derived in ref.~\cite{greensite} for calculating the electromagnetic energy density in three-dimensional $U(1)$ lattice gauge theory, using the polymer formulation of the path integral.

\section{A comparison to a dual Higgs model}

The results of $U(1)$ simulations in the confinement phase verify the dual superconductor picture \cite{haymaker, pl95}: The electric flux between a charge pair is squeezed into a flux tube, encircled by monopole currents acting like a coil. We want to discuss how this scenario can be described by the dual non-compact Abelian Higgs model, in order to establish an exact connection to the dually transformed lattice $U(1)$ theory presented above. The dual Higgs model is written in terms of dual potentials $^*A$ assigned to links on the dual lattice (the $^*$-notation is again used for all fields which are located on the dual lattice) and of a dual Higgs field  $^*\Phi$ which is located on the dual sites. Electric charges in this dual model correspond to monopoles (assigned to three-dimensional cubes on the dual lattice) and can be introduced by the dual of a Dirac string in the same way as in the last section. On the original lattice the world lines of electric charges define the boundary of a continous surface (the Dirac sheet):
\begin{equation}
n = \left\{ \begin{array}{r@{\qquad}l}
			1 & \textrm{for plaquettes on the surface} \\
			0 & \textrm{elsewhere}
			\end{array} \right.
\end{equation}
The field strength ${}^*F$ is given by $g\,{}^*F = g\,d \,{}^*A + 2 \pi {}^* n$, where $d \,{}^*A$ is the exterior derivative of the 1-form $^*A$, and $g=2\pi/e$ is the magnetic coupling. In the London limit the Higgs field $^*\Phi$ is constrained to $^*\Phi=\exp(i {}^*\chi)$ and therefore the coherence length vanishes. Including electric sources the corresponding action reads
\begin{equation} \label{higgs}
S = \tilde{\beta} \sum_{\rm plaquettes} G \left( g\,{d}\, {}^*A + 2 \pi {}^* n\right) \; - \; \tilde{\gamma} \; \sum_{\rm links} \; \cos \left(d\,{}^*\chi - g \,{}^*A\right),
\end{equation}
where $\tilde{\beta}=1/g^2$. The function $G(g\,{}^*F)$ determines the non-compact gauge field action  and approaches $G(g\,{}^*F) \to g^2 \,{}^*F^2/2$ in the continuum limit. The second part of the action is the compact lattice version of the gauge invariant kinetic term of the Higgs field $^*\Phi$. The ratio $\tilde{\beta}/\tilde{\gamma}$ has the meaning of the bare squared London penetration length $\lambda^2$. The Higgs current $^*J_H$, given by
\begin{equation}\label{jh}
g\, {}^*J_H = \frac{\tilde{\gamma}}{\tilde{\beta}} \; \left([{d}\, {}^*\chi] - g\, {}^*A \right),
\end{equation}
is constructed in such a way that the fluxoid quantisation ${}^*F + \lambda^2 {d} {}^*J_H = {}^*m \frac{2\pi}{g} = {}^*m e$ \cite{haymaker} is valid for every single field configuration. The integer two-form $^*m$ characterizes the physical fluxoid string which is allowed to fluctuate in simulations due to the compactness of the phase $\chi$ of the Higgs field. $[{d}\, {}^*\chi]$ in eq.~(\ref{jh}) indicates the reduction of ${d}\, {}^*\chi$ to the interval $(-\pi,\pi]$. Therefore $d [d {}^*\chi]$ may be non-zero for topologically non-trivial field configurations, and $2\pi {}^*m = 2\pi {}^*n + d [d {}^*\chi]$. If the value of $[d {}^*\chi]$ on one dual link flips, i.~e.~the phase difference between neighbouring points exceeds the range $(-\pi,\pi]$, this effects six dual plaquettes corresponding to a closed cube on the original lattice. If such a cube has one plaquette in common with the Dirac sheet ($n=1$), $2\pi {}^*n$ and $d [d {}^*\chi]$ may cancel each other, changing the shape of the physical fluxoid string.

This Higgs model can be regarded as four-dimensional generalisation of the static classical effective model investigated in ref.~\cite{pl95}. It can also serve to extend numerically the predictions from the dual QCD ansatz \cite{baker} or from similar analytical calculations in effective models \cite{andrei}, especially in investigating fluctuations of the fluxoid string. For the $U(1)$ gauge theory, it is even possible to establish an exact connection at the level of the path integral. Let us assume $\tilde{\gamma} \to \infty$ which imposes a new constraint on the fields: Choosing a gauge where the phase $\chi$ of the Higgs field in eq.~(\ref{higgs}) is zero, we realize that the integral over $^*A$ for each dual link reduces to a sum over an integer $^*l$, according to the constraint $^*A=2 \pi {}^*l$. Up to a $\beta$-dependent factor the $\tilde{\gamma} \to \infty$ limit of the partition function of this Higgs model equals the dually transformed path integral (\ref{wje}) of $U(1)$ gauge theory, including two Polyakov loops. This duality property is well-known for the Villain form of the action \cite{froehlich}. In the case of the Wilson action the function $G(g\,{}^*F)$ describing the gauge field action in the corresponding Higgs model is determined by the modified Boltzmann factor in eq.~(\ref{ztp})
\begin{equation}
\exp\left[-\tilde{\beta} G\left(g\,{}^*F\right)\right] = e^{-\beta} I_{\parallel g\,{}^*F \parallel}(\beta), 
\end{equation}
and $\tilde{\beta}=1/(4\pi^2\beta)$.

Especially interesting for us is the interpretation of expectation values in $U(1)$ in terms of the dual Higgs model. Let us first regard the field strength in $U(1)$ derived in eq.~(\ref{sin}): It also agrees with the field stength of the dual Higgs model in the $\tilde{\gamma} \to \infty$ limit which reads
\begin{equation}
g {}^*F \to 2\pi d \,{}^*l + 2\pi ^*n  \; \Rightarrow \; ^*F \to e (d \,{}^*l + {}^*n) = \frac{1}{\sqrt{\beta}} {}^*k.
\end{equation}
This means that the field strength expectation values and due to the validity of the extended Maxwell equations also the magnetic currents in $U(1)$ gauge theory agree with the corresponding fields and currents in the dual Higgs model in the limit of both zero coherence length and zero penetration length. From this point of view a $U(1)$ flux tube looks rather like an infinitely thin fluctuating string than like a flux tube in a classical dual superconductor. 

This correspondence to the Higgs model does not hold for energy densities. The dually transformed expectation value of the squared field strength, eq.~(\ref{cos}), obviously disagrees with the corresponding result in the $\tilde{\gamma} \to \infty$ limit of the dual Higgs model. This is not astonishing, since the expectation values of the squared fields are connected to the vacuum fluctuations which show the opposite behaviour in the dual theory, where the confinement phase is the weakly coupled phase and fluctuations become more important with increasing $\beta$. Hence, a direct interpretation of the $U(1)$ energy density in the dual Higgs model is not possible, which makes it difficult to distinguish by analytical considerations between dual type-I and type-II superconductivity. 

\section{Simulating the dually transformed $U(1)$ theory}

As has been pointed out above, one has to be careful when comparing $U(1)$ lattice gauge theory to a dual Higgs model on a ``microscopic'' level. Nevertheless, simulating the dual $U(1)$ gauge theory gives us the possibility to calculate any expectation value in the presence of static charges with significantly higher precision. This has several reasons: The confinement phase is the weakly coupled one in the dual theory, therefore we have less quantum fluctuations in the region of interest. Simulating the dually transformed path integral (\ref{sin}) resp.~(\ref{cos}) is also much faster because of the integer valued gauge fields. Most important, however, is the fact that it is not necessary to project the charge--anticharge state out of the vacuum: Charge pairs with arbitrary distance can be simulated with equal accuracy, as well as multiply charged systems \cite{stl,pub3}.

The updating procedure itself is very simple: As already discussed in the previous section, the dual Dirac string sheet $^*n$ connecting the two Polyakov lines is held fixed, while the integer gauge field $^*l$, located on the dual links, is changed by $\pm 1$ during the updates. We have to mention that this is essentially the same procedure as the ``change-a-cube'' algorithm used in ref.~\cite{greensite}. The only difference for the three-dimensional model used there is that the dual potential is located on the dual sites instead of the dual links. In ref.~\cite{greensite} the electromagnetic energy distribution was calculated to extract the string tension in a range of weaker couplings as total field energy per length. However, this quantity must not be confused with the string tension calculated from usual correlation functions.

Recently, a detailed analysis of simulating dual $U(1)$ lattice gauge theory in three dimensions was performed in ref.~\cite{coyle}. The authors found that at weak coupling the advantages are outweighed by large autocorrelation times as the dual model becomes disordered. We also observed the increase of the autocorrelation time and correspondingly of the error for increasing values of $\beta$ -- with a peak at the phase transition of the four-dimensional theory. In the deconfinement phase there are large fluctuations, but the autocorrelation time is smaller than around the phase transition. Since our goal is the investigation of the confinement mechanism, we take the main advantage from simulations in the confinement phase, anyhow. In this context we want to mention that we used a duality transformation exactly valid for the whole $\beta$ range, not only in the weak coupling limit as in ref.~\cite{coyle}. Further, as stated above, the power of dual simulations lies in the inclusion of the sources in the action, which already has been realized in ref.~\cite{greensite,caselle}.  

For each charge distance $d$ a separate simulation has to be performed, and of course for the subtraction of the vacuum expectation value in (\ref{correlation}) a distinct run is needed. Therefore one gets a good estimate of the reliability of the errorbars. For smaller distances on a $8^3 \times 4$ lattice an accurate comparison was performed between expectation values from ordinary and from dual simulations \cite{stl}. The only difference in the observed results is the effect of the periodic boundary conditions which is absent in the dual model: There can be no electric flux over the period per construction. For $d \geq 3a$ the dual simulation is clearly more efficient, because in ordinary simulations the signal decreases exponentially with the charge distance $d$, while in the dual simulation it stays constant (the necessary amount of computer time still depends of course on the lattice size used).

For the updating procedure we use a standard Metropolis algorithm, which has proven sufficient for the investigated range of couplings, although a nonlocal cluster algorithm would certainly lead to a further increase in performance. Our typical measurements are taken from $5\cdot10^5 - 2\cdot10^6$ configurations, where each 10th configuration is evaluated. For determining the errorbars we took blocks of 100 evaluated configurations. For the results shown in the next sections we used the coupling $\beta=0.96$, where the system is clearly in the confinement phase. On the other hand the fluctuations in the dual theory are already large enough to produce a flux tube of the desired transverse extent. If one likes to compare with the QCD flux tube, we can require that our string tension equals the physical value, which gives a lattice spacing $a$ of roughly $0.25 fm$.

\section{The profile of flux tubes between static charges}

We present now our numerical results for the electric field and magnetic current distribution in the presence of a static charge pair. The great advantage compared to the results reported in ref.~\cite{pl95} is that we are now able to investigate much larger charge distances and much lower temperatures.

\begin{figure}
\centerline{\epsfbox{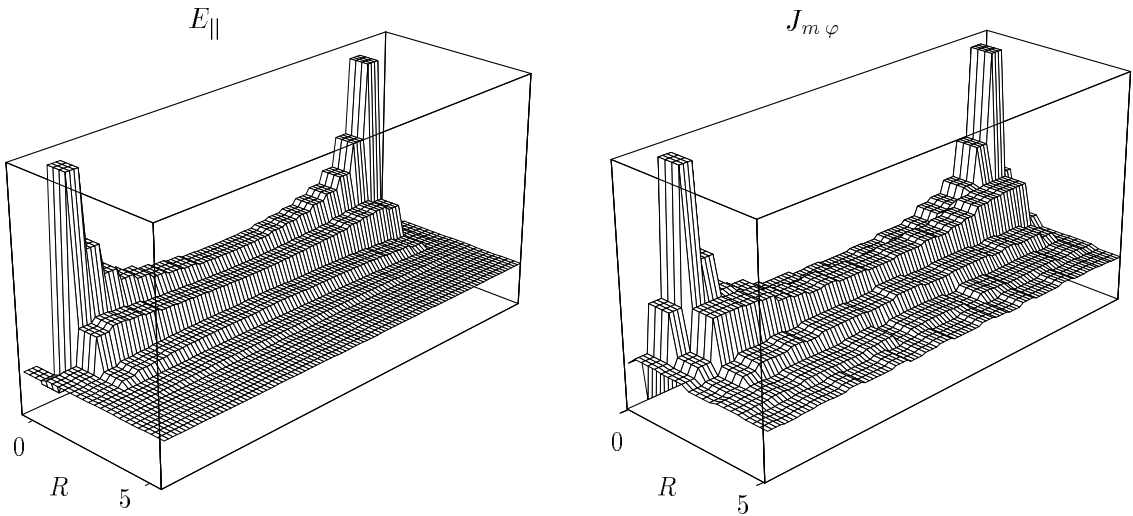}}
\caption{\label{plot3d}The flux tube between static charges for a distance of 22 lattice spacings. The left plot shows the parallel component of the electric field, the right plot shows the azimuthal component of the magnetic current, both in a half plane through the $Q\bar{Q}$-axis. Note that in the symmetry plane between charges the magnetic current takes its maximum value at a transverse distance of $1.5 a$, not at $R=0.5a$.}
\end{figure}

Before discussing the temperature dependence of the flux tube profile, let us comment on the necessary spatial size of the lattice. There is the simple condition that the spatial extent of the lattice has to be greater than the extent of the considered flux tube both in longitudinal and in transverse direction in order to prevent strong finite size effects. For our simulations we used a spatial extent of $8^3$ for charge distances $d=1a - 5a$, $12^3$ for simulating $d=5a - 8a$, and $16^3$ for $d=8a - 12a$. Further we performed some simulations on larger lattices; the largest investigated charge distance was $d=22a$ (on a $26^3 \times 32$ lattice). Results for this distance are displayed in a 3d-plot in fig.~\ref{plot3d}. We focus on the longitudinal component of the electric field strength and on the azimuthal component of the magnetic current which encircles the flux tube like a coil. It can already be seen in these plots that a flux tube of constant thickness is not observed even for this large charge distance. The increase of the flux tube width with the distance between charges may be explained by roughening effects and will be discussed in section \ref{wsq}. The right figure demonstrates that the signal for the magnetic current drops in the middle between charges, its radial dependence in the symmetry plane exhibits a maximum at finite $R$. This behaviour is interesting for the dual superconductor interpretation as will be discussed in the next section.

\begin{figure}
\centerline{\epsfbox{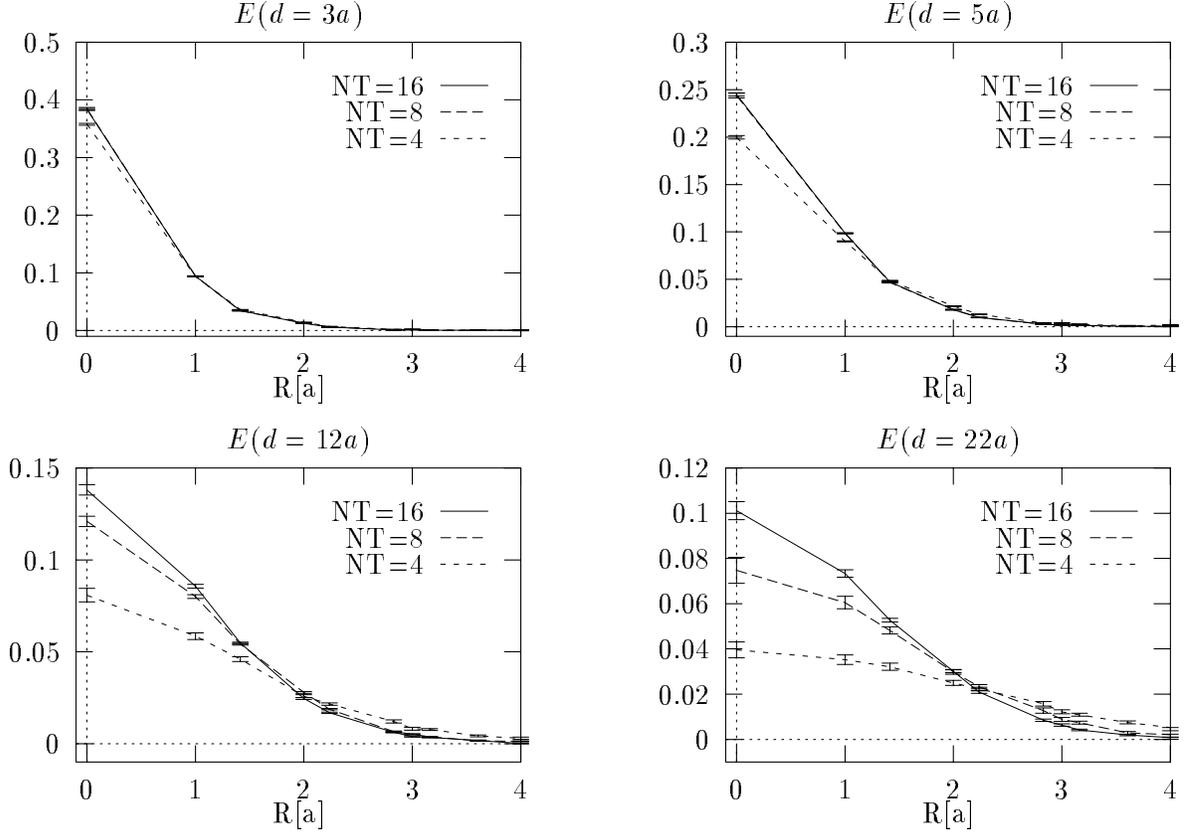}}
\caption{\label{e_nt}The temperature dependence of the electric field profile in the symmetry plane between charges for various charge distances $d$: If the time extent NT of the lattice (corresponding to the inverse temperature) is smaller than the charge distance, the width of the flux tube is strongly temperature dependent.}
\end{figure}

\begin{figure}
\centerline{\epsfbox{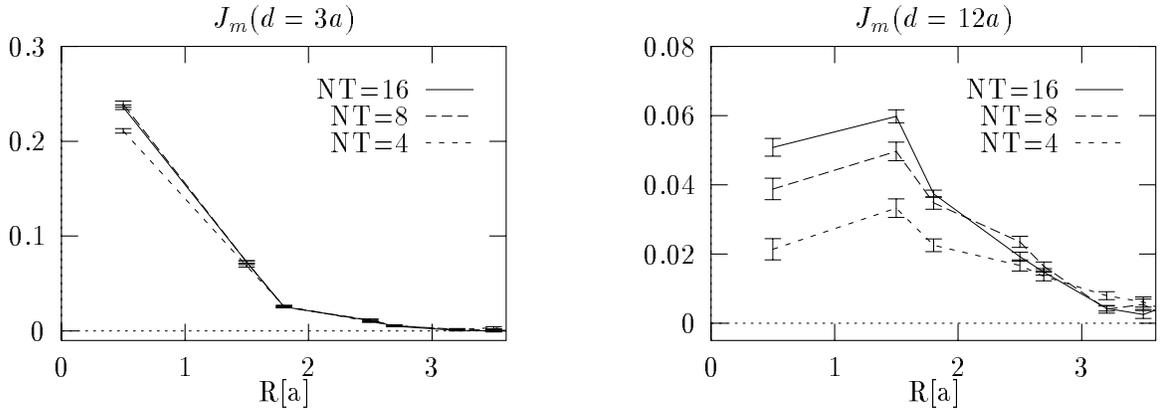}}
\caption{\label{j_nt}The temperature dependence of the magnetic current profile in the symmetry plane between charges for charge distances $d=3a$ and $d=12a$: If NT is smaller than $d$, the signal for the current gets significantly lower which means a widening of the flux tube. For $d=12a$ the maximum current is observed at the transverse distance $R=1.5a$, like in the $d=22a$ case shown in fig.~\ref{plot3d}.}
\end{figure}

In all our simulations we represent static charges by Polyakov loops and therefore have to deal with finite temperature effects due to the finite time extent of the lattice. It is therefore important to separate the occuring thermal fluctuations from quantum fluctuations which are essential for the extended flux tube. This is achieved by analyzing the dependence of the results on the time extent NT, as is shown in fig.~\ref{e_nt} for the electric field profile and in fig.~\ref{j_nt} for the magnetic current profile in the symmetry plane between charges for various charge distances $d$. From the results we conclude that if the time extent NT of the lattice exceeds the charge distance $d$ finite temperature effects can safely be neglected. For smaller NT, however, we realize a significant widening of the flux tube, connected with a decreasing magnetic current. From the point of view of the dual simulation, this widening can easily be explained. Quantum fluctuations of the world sheet of the string correspond to cube-like excitations on the original lattice. At finite temperature there also appear time-independent excitations which wrap around the lattice in time direction and therefore increase the effective flux tube width. These excitations vanish if NT is large compared to the length of the flux tube.

After the discussion of the techniques used for the simulation of long flux tubes and the dependence of the results on the temperature, we will now turn again to the dual superconductor interpretation.

\section{A closer look at the dual London equation}\label{mod}

In the presence of a pair of static charges, the classical solution of the Higgs model (\ref{higgs}) yields an electric field and a magnetic current distribution according to
{\begin{eqnarray}
\label{ampere}
{\rm rot}\vec{E}(\vec{r}) &=& -\vec{J}_m(\vec{r}),\\
\label{fluxoid}
\vec{E}(\vec{r}) &=& \lambda^2 {\rm rot}\vec{J}_m(\vec{r}) + \vec{\cal E}(\vec{r}),
\end{eqnarray}}where the (straight) string $\vec{\cal E}(\vec{r})$ connects the point charges, carrying a quantum of electric flux $e$. Integrating eq.~(\ref{fluxoid}), a generalisation of the London equation, over a surface intersecting the string gives the fluxoid quantisation. For an infinitely long string eq.~(\ref{ampere}) and eq.~(\ref{fluxoid}) yield an electric field profile of the form ${E}(R) = (e / 2\pi\lambda^2) K_0(R/\lambda)$.

These classical equations were used in ref.~\cite{haymaker} for comparison with the results of $U(1)$ lattice simulations. The free parameter $\lambda$ (the London penetration length) could be fitted by comparing $\vec{E}$ and ${\rm rot}\vec{J}_m$ off axis, and also the fluxoid quantisation was verified. An extension of this analysis was performed in ref.~\cite{pl95} by including fluctuations of the string $\vec{\cal E}(\vec{r})$ in the above model. Further the effective model was solved numerically on a lattice of same size, to avoid being misled by lattice artifacts. Due to the exponential falloff of the signal, however, the $U(1)$ data analyzed so far contain only charge distances up to $d=3a$. We will now consider the data obtained in the dual simulations, in order to see if the above interpretation holds for longer flux tubes.

\begin{figure}
\centerline{\epsfbox{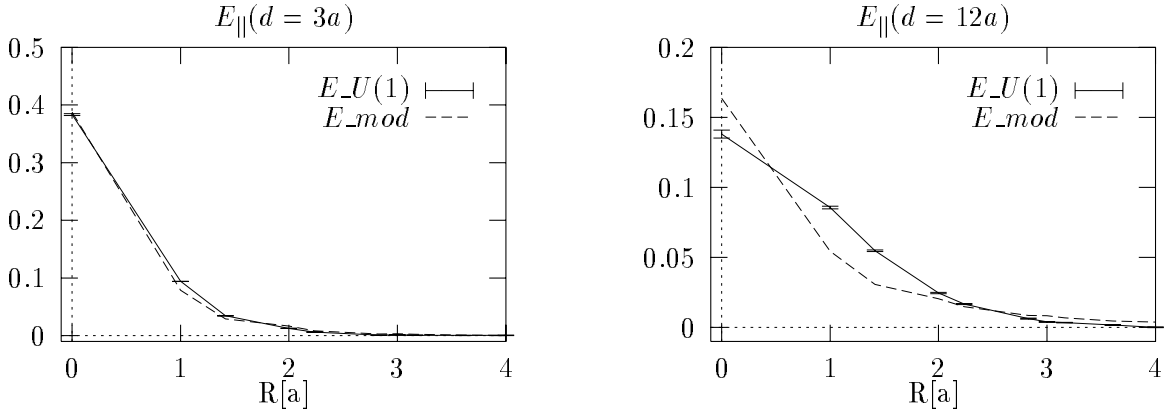}}
\caption{\label{e_mod}Comparison of the $U(1)$ electric field profile with the numerical solution of Maxwell and London equation for a straight fluxoid string. The free parameter $\lambda$ is fitted. For small charge distances ($d=3a$, left figure, $\lambda_{fit}=0.75a$) good agreement can be achieved; for larger distances ($d=12a$, right figure, $\lambda_{fit}=1.50a$) it is not possible to describe the $U(1)$ flux tube behaviour.}
\end{figure}

In the classical model with a straight string $\vec{\cal E}(\vec{r})$ one would expect that
\begin{itemize}
\item{the flux tube width as a function of $d$ soon approaches a constant, this means that the fitted parameter $\lambda$ should not depend on $d$,}
\item{$\lambda^2$ is constant everywhere off axis,}
\item{the azimuthal magnetic current encircling the axis should exhibit a maximum in the symmetry plane between the charges, and not near the charges. The radial dependence should exhibit an exponential falloff.}
\end{itemize}
It was shown that these requirements are not even fulfilled for a charge distance $d=3a$ \cite{pl95}, nevertheless the flux tube profile in the symmetry plane may be described well by the effective model with a static string \cite{haymaker}. Considering larger charge distances confirms the discrepancy to the items mentioned above. In a comparison of electric field profiles for $d=3a$ and $d=12a$ (see fig.~\ref{e_mod}) it can be seen that the fitted values for $\lambda$ not only differ very much, but it is already impossible to describe the field distribution for larger charge distances by the numerical solution of coupled Maxwell and London equation.

\begin{figure}
\centerline{\epsfbox{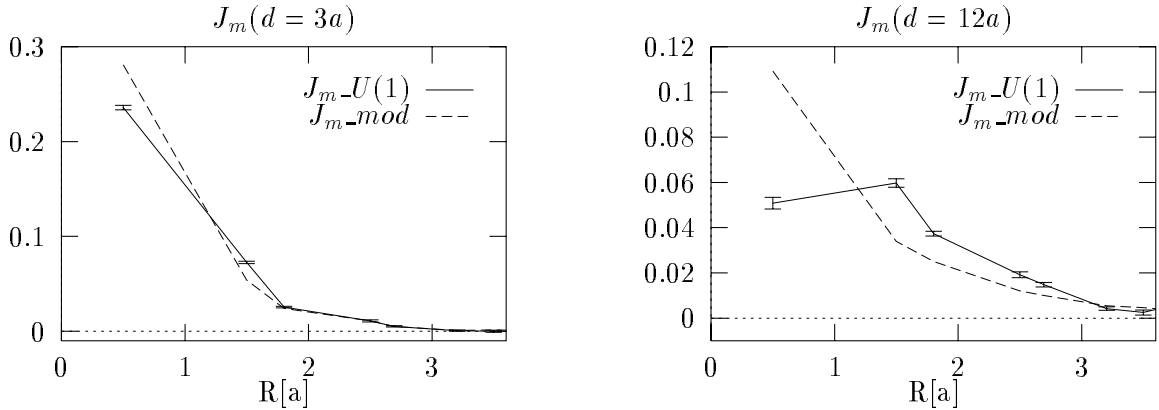}}
\caption{\label{jm_mod}Comparison of the $U(1)$ magnetic current profile with the numerical solution of Maxwell and London equation for a straight fluxoid string ($\lambda$-values chosen as in fig.~\ref{e_mod}). The agreement is reasonable for small charge distances ($d=3a$, left figure), for large distances the behaviour of the current distribution cannot be described by the solution of the London model ($d=12a$, right figure).}
\end{figure}

For the magnetic current distribution we already see in fig.~\ref{plot3d} and in fig.~\ref{j_nt} that its behaviour cannot be explained in the above effective model. Fig.~\ref{jm_mod} shows the corresponding classical solutions, again for $d=3a$ and $d=12a$. For larger distances, there is not even qualitative agreement: The azimuthal component of the magnetic current as a function of the transverse distance first increases and takes its maximum value at $R=1.5a$. 

This phenomenon together with the other described observations may be explained within the dual superconductor picture by two different approaches: First, it can be regarded as evidence for the importance of string fluctuations. This way was chosen in ref.~\cite{pl95}, where the intrinsic thickness of the flux tube $\lambda$ turned out to be much smaller than the effective flux tube width which increases with charge distance. The extreme scenario ($\lambda \to 0$) leads us again to the exact correspondence of $U(1)$ to the $\tilde{\gamma} \to \infty$ limit of a dual Higgs model.

The second possibility for arguing is to stay in the classical model of a dual superconductor, but leave the London (extreme type II) limit described so far.
A strong suppression of the Higgs condensate in the region of electric fields may also lead to the behaviour of the current observed in fig.~\ref{jm_mod}. A similar current distribution measured in Abelian projection of $SU(2)$ was interpreted in this way in ref.~\cite{haymsu2}. Hence, one may conclude that $U(1)$ lattice gauge theory in this respect behaves like a dual type-I superconductor. Stronger evidence for such a statement is expected of course from an investigation of the interaction between flux tubes (which also can be performed very efficiently in a dual simulation). Indeed preliminary results \cite{pub3} show an attraction between $U(1)$ flux tubes for the regarded value of $\beta$.

We may conclude that interpreting the flux tube results in terms of the dual superconductor picture still makes sense, but $U(1)$ lattice gauge theory certainly cannot be regarded as a classical dual superconductor in the London limit if one examines larger charge distances. On the other hand the effective string description of flux tubes represents an independent feature of the lattice gauge system, and from the point of view of dual simulations this looks somehow easier to argue. We will use this approach in the next section for a closer look at the flux tube width in dependence of the charge distance.

\section{The flux tube width and the effective string description}\label{wsq}

In ref.~\cite{luescher} the string model approximation to Wilson loop correlation functions was used to study the behaviour of the flux tube width for large charge distances. This model assumes a very thin bare flux tube subjected to quantum mechanical fluctuations. A convenient measure for the transverse extent of the flux tube is
the squared width
\begin{equation}
w^2=\frac{\int r^2_{\perp} {\epsilon}(r_{\perp}) d^2r_{\perp}}{\int {\epsilon}(r_{\perp}) d^2r_{\perp}},
\end{equation}
where {$\epsilon$} represents a quantity characteristic of the flux tube, for example the (chromo-)electric field energy density. It was found in ref.~\cite{luescher} that the squared width $w^2$ diverges logarithmically for $d \to \infty$,
\begin{equation}\label{wlog}
w^2 = w_0^2 \, \ln (d/d_c)
\end{equation}
with some constants $w_0^2$ and $d_c$. A numerical test of this hypothesis could not be performed up to now for $U(1)$ lattice gauge theory. In the three-dimensional $Z_2$ gauge model this prediction was verified in ref.~\cite{caselle}, where the authors also exploited the duality relation to a three-dimensional Ising model and combined their data with high precision data on the interface physics of the three-dimensional Ising model. They also achieved agreement with the prediction
\begin{equation}\label{w0sq}
w_0^2=\frac{1}{2\pi\sigma}
\end{equation}
(where $\sigma$ is the string tension) for the two-dimensional free Gaussian model. There is the conjecture that this behaviour of the flux tube width should be observed in all gauge theories. 

\begin{figure}
\centerline{\epsfbox{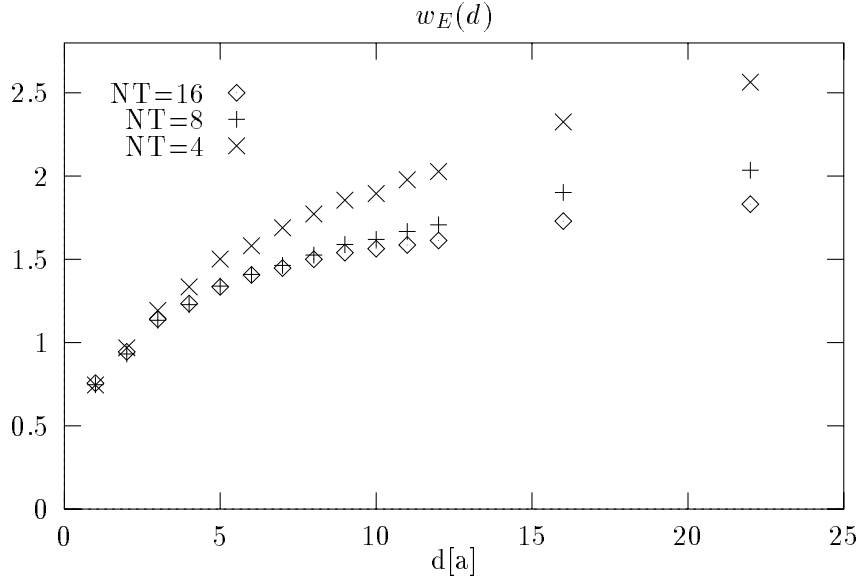}}
\caption{\label{w_nt}The thickness of the flux tube $w_E$ (calculated from the profile of the electric field) in the symmetry plane between charges as a function of the charge distance $d$ for various time extents NT of the lattice. If $d$ exceeds the time extent NT, finite temperature effects become important and the flux tube significantly widens.}
\end{figure}

\begin{figure}
\centerline{\epsfbox{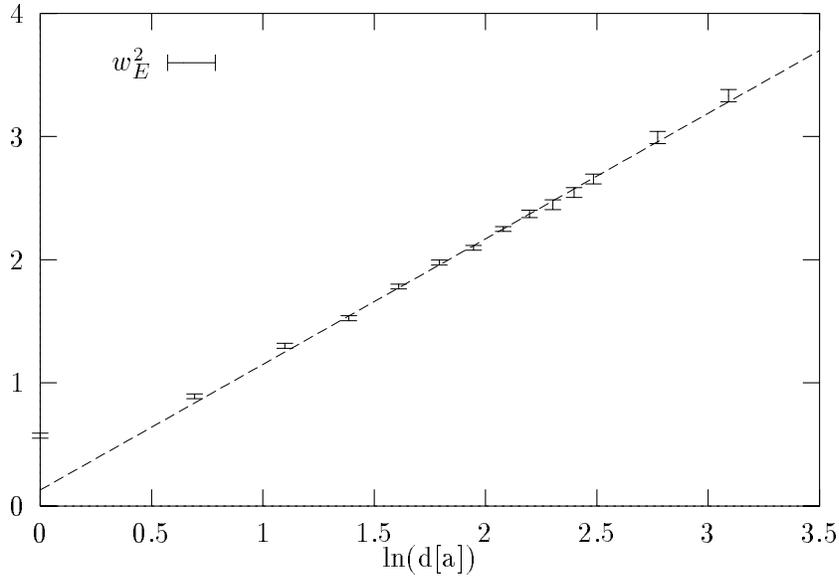}}
\caption{\label{cas}The squared thickness of the flux tube $w^2_E$ (for NT=16) is plotted against the logarithm of the charge distance $d$. The dashed line shows a fit of the form $w^2 = w_0^2 \, \ln (d/d_c)$ for charge distances $d \geq 3$, with $w_0^2=1.02(2)$, $d_c=0.88(4)$ and $\chi^2 < 1$.}
\end{figure}

It is therefore interesting to test this prediction in dual $U(1)$ simulations where large flux tubes become accessible. We determine the flux tube width by means of the parallel component of the electric field strength $E_{\|}$. Our defintion on the lattice reads:
\begin{equation}
w_E^2=\frac{\sum_i R^2_i E_{\|}(R_i)}{\sum_i E_{\|}(R_i)}.
\end{equation}
This slightly reduces the estimated error compared to the definition via the electric field energy, in all other aspects the results roughly agree. The sum extends over a transverse plane of the lattice with $R_i < R_{max}$: For $d \leq 12a$ we evaluate the sum until a transverse distance $R_{max}=4a$, for larger distances we use $R_{max}=5a$. In fig.~\ref{w_nt} one can see again the influence of finite temperature effects on the flux tube width. The calculation with time extent 16 can be taken as acceptable approximation to the zero temperature case for the regarded charge distances and is analyzed according to the prediction (\ref{wlog}) in fig.~\ref{cas} in analogy to ref.~\cite{caselle}. The logarithmic increase of the flux tube width is realized very well for $U(1)$ lattice gauge theory. The constant $w^2_0$, however, disagrees with expr.~(\ref{w0sq}) by roughly a factor 2.

The above data demonstrate that the thickness of $U(1)$ flux tubes does not saturate for large charge distances, as expected in a naive formulation of the dual superconductor picture. It seems to diverge logarithmically, as predicted by the effective string description. As can be seen from fig.~\ref{cas} this behaviour sets in already at $d=2a$, which corresponds to a physical length of around $0.5fm$ in agreement with ref.~\cite{caselle}. 

\section{Energy from the dual simulation}

In section 2 we have discussed the duality transformation for the expectation value of $\beta \cos d \theta_{\Box_0}$ in the presence of two Polyakov loops -- see eq.~(\ref{cos}) -- which gives the squared field strength on a specified plaquette $\Box_0$. This allows for the determination of the Euclidean action density $E^2+B^2$ and energy density $E^2-B^2$. As has been pointed out, the squared field strength calculated in the dual simulation differs from the squared field strength of the $\tilde{\gamma} \to \infty$ Higgs model. The contributions of all plaquettes can be summed up and give the total action respective energy of the system. In ordinary lattice simulations this is a very difficult task, since the difference between the squared field strength of the pure $U(1)$ vacuum and the charge--anticharge state is small compared to the vacuum expectation value itself. Therefore it cannot be resolved for larger charge distances. Further, concerning the energy density there are large cancellations between electric and magnetic field contributions.

\begin{figure}
\centerline{\epsfbox{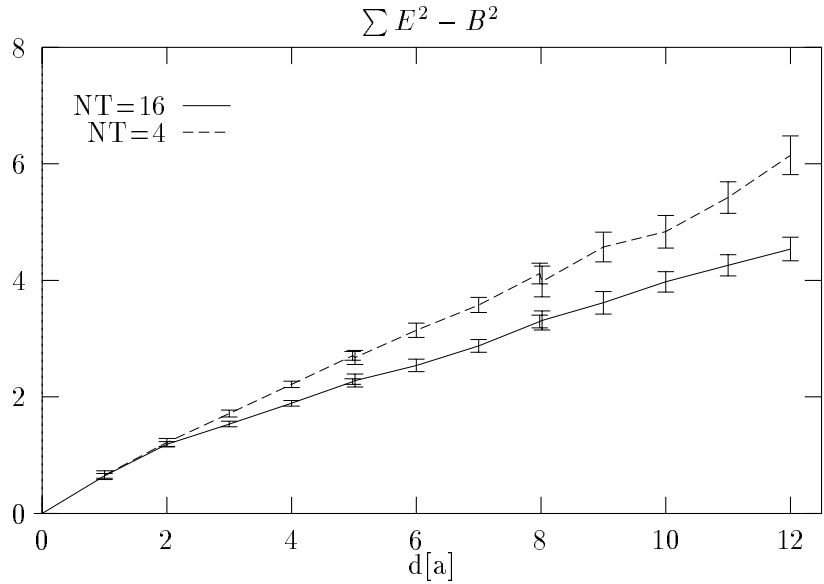}}
\caption{\label{en_16_4}The total field energy as a function of the charge distance $d$ for NT=16 and NT=4. For higher temperature the periodic excitations in time direction mainly contribute to $E^2$ and therefore increase the total energy.}
\end{figure}

In the dual simulation of $U(1)$ lattice gauge theory it is possible to determine the total energy of the electromagnetic fields as a function of the charge distance. The influence of finite temperature effects is analyzed in fig.~\ref{en_16_4}. The thermal fluctuations increase the total energy. This is again due to the periodic excitations in time direction which contribute to the electric part of the energy only.

The total field energy shown in fig.~\ref{en_16_4} qualitatively behaves like the free energy for a charge pair (as determined in usual lattice simulations), i.e.~up to a constant like a Coulomb and a linear term. However, as already realized in ref.~\cite{greensite}, a quantitative comparison shows that the integral over the energy density differs from the potential. We mention here that it is also possible to determine the free energy via dual simulations; this opens the opportunity for the application of lattice sum rules \cite{michael}. Such calculations are in progress and will be reported elsewhere.

\begin{figure}
\centerline{\epsfbox{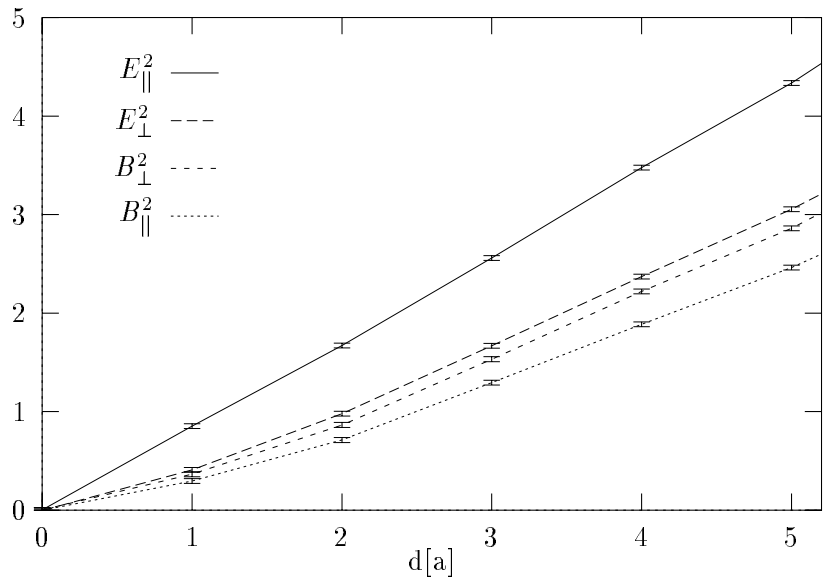}}
\caption{\label{fsq}The various contributions to the total energy (for NT=16). The parallel component of the electric field increases almost linearly for all distances, the parallel magnetic component develops its full slope after $d=3a$ only.}
\end{figure}

It is also interesting to look at the individual contributions to the total field energy. For $\beta \to 0$ there are no string fluctuations, and therefore only the parallel component of the electric field contributes. If we consider single cube-like excitations in transverse directions, we get also contributions from transverse electric and transverse magnetic fields. In this case they cancel each other exactly and therefore do not change the total energy. For stronger fluctuations, i.~e.~increasing roughening of the string, these components differ, and also the parallel magnetic component gets nonzero. The results for $\beta=0.96$ are plotted in fig.~\ref{fsq}. One can argue that for charge distances $d < 3a$ the string fluctuations are not fully developped. The ``Coulomb'' behaviour of the total field energy observed in fig.~\ref{en_16_4} is mainly due to the parallel component of the magnetic field which gives the smallest contribution to the linear increase of the energy. This demonstrates that the behaviour of the flux tube cannot be explained easily within a classical static model.

\section{Conclusions}

Our motivation was a detailed quantitative analysis of the dual superconductor picture in $U(1)$ lattice gauge theory. By considering the duality transformation of correlation functions we improved the understanding of the behaviour of expectation values of fields and currents in the presence of a charge pair, as well as the significance of string fluctuations. Moreover, simulating the dually transformed theory provided numerical flux tube data for low temperatures and large charge distances not obtained before in four-dimensional $U(1)$ lattice gauge theory.

We found that finite temperature effects may be neglected for time extents exceeding the charge distance, while for smaller time extents the flux tube widens significantly. Further we examined the electric field and magnetic current distribution and found that its description by the classical solution of Maxwell and London equation fails for large charge distances. The observed effects may be explained by fluctuations of the flux tube, on the other hand the current profile resembles that of a dual type-I superconductor. Roughening effects which become important for large distances are described well by the effective string picture of confinement which predicts a logarithmic increase of the flux tube width. Finally, the analysis of the total field energy revealed the responsibility of the parallel magnetic field for the Coulomb behaviour at small distances.

At this time, no final statement should be made which effective model is really able to describe strongly coupled $U(1)$ gauge theory. Certain aspects of the flux tube may be explained by a ``classical'' dual superconductor picture, while others cannot. Further studies considering periodically closed flux tubes (torelons) are in progress and promise to answer some of the remaining questions.

\section*{Acknowledgments}

We want to thank Richard W.~Haymaker for interesting discussions on the description of confinement in $U(1)$ lattice gauge theory by effective models. Further we thank Jeff Greensite for making us aware of an article (formerly not known to us), where the energy density in three-dimensional $U(1)$ was calculated in an analogous way. Finally we acknowledge critical comments from Andrei Ivanov and \v{S}tefan Olejn{\'\i}k.


\end{document}